\documentclass[showpacs,amsmath,superscriptaddress,reprint,aps]{revtex4-2}

\usepackage{graphicx}
\usepackage{hyperref}

\begin{document}

\title
{Quantization of mode shifts in nanocavities integrated with atomically thin sheets}
\author{N.~Fang}
\email{nan.fang@riken.jp}
\affiliation{Nanoscale Quantum Photonics Laboratory, RIKEN Cluster for Pioneering Research, Saitama 351-0198, Japan}
\author{D.~Yamashita}
\email{daiki.yamashita@riken.jp}
\affiliation{Quantum Optoelectronics Research Team, RIKEN Center for Advanced Photonics, Saitama 351-0198, Japan}
\author{S.~Fujii}
\affiliation{Quantum Optoelectronics Research Team, RIKEN Center for Advanced Photonics, Saitama 351-0198, Japan}
\author{K.~Otsuka}
\affiliation{Nanoscale Quantum Photonics Laboratory, RIKEN Cluster for Pioneering Research, Saitama 351-0198, Japan}
\affiliation{Department of Mechanical Engineering, The University of Tokyo, Tokyo 113-8656, Japan}
\author{T.~Taniguchi}
\affiliation{International Center for Materials Nanoarchitectonics, National Institute for Materials Science, Ibaraki 305-0044, Japan}
\author{K.~Watanabe}
\affiliation{Research Center for Functional Materials, National Institute for Materials Science, Ibaraki 305-0044, Japan}
\author{K.~Nagashio}
\affiliation{Department of Materials Engineering, The University of Tokyo, Tokyo 113-8656, Japan}
\author{Y.~K.~Kato}
\email{yuichiro.kato@riken.jp}
\affiliation{Nanoscale Quantum Photonics Laboratory, RIKEN Cluster for Pioneering Research, Saitama 351-0198, Japan}\affiliation{Quantum Optoelectronics Research Team, RIKEN Center for Advanced Photonics, Saitama 351-0198, Japan}

\begin{abstract}
The unique optical properties of two-dimensional layered materials are attractive for achieving increased functionality in integrated photonics. Owing to the van der Waals nature, these materials are ideal for integrating with nanoscale photonic structures. Here we report on carefully designed air-mode silicon photonic crystal nanobeam cavities for efficient control through two-dimensional materials. By systematically investigating various types and thickness of two-dimensional materials, we are able to show that enhanced responsivity allows for giant shifts of the resonant wavelength. 
With atomically precise thickness over a macroscopic area, few-layer flakes give rise to quantization of the mode shifts. We extract the dielectric constant of the flakes and find that it is independent of the layer number down to a monolayer. Flexible reconfiguration of a cavity is demonstrated by stacking and removing ultrathin flakes. With an unconventional cavity design, our results open up new possibilities for photonic devices integrated with two-dimensional materials.
\end{abstract}

\maketitle
\section{Introduction}
Two-dimensional (2D) layered materials such as graphene, transition metal dichalcogenides (TMDs), and hexagonal boron nitride (h-BN) have attracted considerable attention due to their exotic physical properties and potential for diverse applications. Monolayer graphene exhibits the quantum Hall effect even at room temperature~\cite{Novoselov:2007}, while ultra-thin TMD based field-effect transistors are promising for continuation of the Moore’s  Law ~\cite{Desai:2016}.  Moir\'e  patterns  can  be  created  by  layers  with  twisted  angles,  which  has led to the observations of superconductivity in bilayer graphene~\cite{Cao:2018} and ferroelectricity in bilayer h-BN~\cite{Yasuda:2021}. Many TMDs are found to become direct bandgap semiconductors at the monolayer limit, making them the thinnest optical emitters~\cite{Mak:2010}. Furthermore, rich optical phenomena exist in these atomically thin flakes, such as spin-valley coupling~\cite{Xiao:2012}, Moir\'e excitons~\cite{Jin:2019}, and layer number dependent nonlinearity~\cite{Saeynaetjoki:2017}.

To further utilize the optical properties of TMDs, microcavities can be used to enhance the light-matter interaction. By forming TMD/cavity hybrids, nanolasers have been demonstrated~\cite{Li:2017,Wu:2015,Shang:2017} and second-order nonlinear response has significantly improved~\cite{Busschaert:2020,Fryett:2016}. Evidence for strong coupling and formation of microcavity polaritons have also been reported~\cite{Liu:2015nature}. Compared to the successful use of TMDs as an active material within microcavities, it is a challenge to control the cavity modes with 2D materials. Because of the atomically thin nature, cavity modes are barely affected and the resonant wavelength can only be tuned slightly~\cite{Zhao:2018,Majumdar:2013,He:2021,Datta:2020,Fryett:2018}. With atomically precise thickness over macroscopic areas and extensive compatibility offered by van der Waals interface, photonic devices controlled through single atomic layers would present a new direction in nanotechnology. Cavity structures designed specifically for enhanced sensitivity to ultra-thin materials is required to unravel the full functionality and capability of the TMD/cavity hybrids. 

Here we substantially boost the responsiveness of microcavities to 2D materials by using specially designed photonic crystal nanobeam structures. Various 2D material types and thickness dependence are investigated, and we find that it is possible to achieve giant wavelength shifts of the cavity mode. By precisely controlling tungsten diselenide (WSe$_2$) thickness down to a monolayer, we observe quantized mode shifts of the 2D/cavity hybrids. The dielectric constant in the atomically thin limit can be extracted by comparing to finite-difference time-domain (FDTD) calculations. Stacking and removal of WSe$_2$ flakes are also performed, demonstrating the flexible reconfiguration capability of the 2D/cavity hybrids.

\section{Results and discussion}
\paragraph*{Design of 2D/cavity hybrids.}

\begin{figure*}
\includegraphics{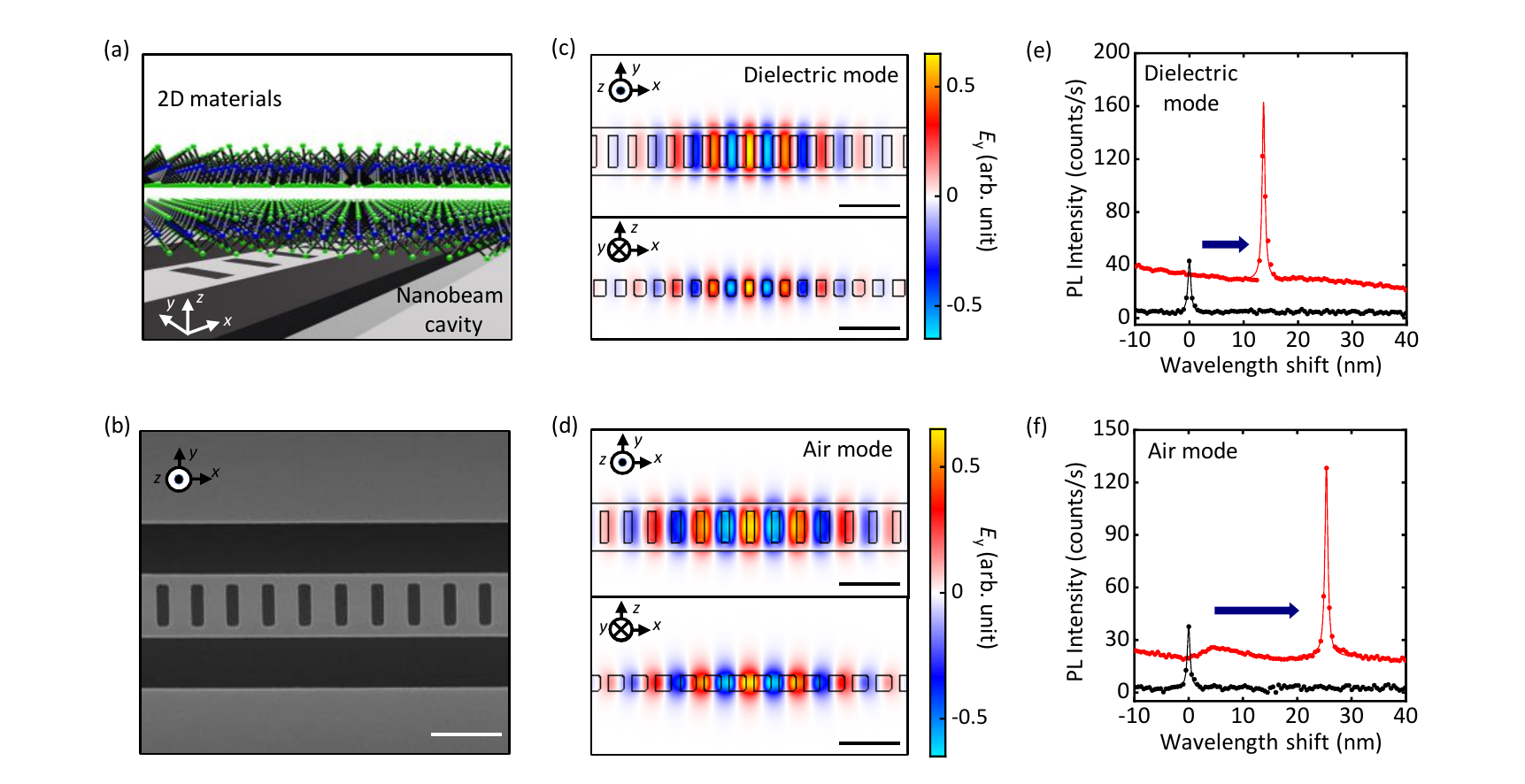}\
\caption{
\label{Fig1} Design of PC nanobeam cavities. (a) A schematic of a 2D/cavity hybrid. (b) A top-view SEM image of a bare nanobeam cavity. Spatial distribution of $E_y$ for the fundamental cavity modes of (c) dielectric-mode and (d) air-mode structures with top views (top) and cross-section views (bottom). PL spectra of the fundamental mode for (e) a dielectric-mode nanobeam and (f) an air-mode nanobeam with (red) and without (black) a 9.0-nm-thick~WSe$_2$ flake on top. The excitation power is 300~$\mu$W and the excitation wavelength is 780~nm. Spectra are plotted with respect to the resonant wavelengths before the transfer of WSe$_2$ flakes, which are 1405.6 nm and 1477.4 nm for (e) and (f), respectively. We note that the initial resonant wavelength in the present range barely affects the shift values in the FDTD simulation. Dots are data and solid lines are Lorentzian peak fits. Arrows indicate the shift directions. Scale bars in (b--d) are 1~$\mu$m. 
}
\end{figure*}

We begin by designing cavities suited for interaction with 2D materials. Most microcavities, such as microsphere resonators~\cite{Righini:2011,Hall:2015}, are designed based on refractive index contrast between two media, and the modes are confined in the high-index medium using total internal reflection. The evanescent fields are typically small outside of the cavity, and coupling to 2D materials would be weak. Reducing the size of the cavity would force the mode to extend outside, and therefore smaller mode volumes are in general favorable for enhanced evanescent fields. For example, toroidal microcavities having small diameters show more of the modes occupying the space external to the cavity~\cite{Min:2007}.

We propose to use photonic crystal (PC) nanobeam cavities with ultra-small mode volumes~\cite{Foresi:1997}, consisting of an array of air holes on a silicon waveguide as shown in Fig.~\ref{Fig1}a, b. Bragg reflectors are formed by periodic arrays of holes, and localized modes are formed by modulating the periodicity at the cavity center to break the Bragg condition. The nanobeam cavities feature ultra-small mode volumes on the order of $10^{-2} \lambda^3$ where $\lambda$ is the cavity wavelength. In addition, the air holes act to reduce the average refractive index, weakening the confinement and increasing the evanescent fields.  

We further utilize the characteristics of the photonic bands to engineer the cavity modes for enhancing the interaction with 2D materials. The periodic modulation of the refractive index gives rise to a photonic bandgap, separating the lower frequency dielectric band from the higher frequency air band~\cite{Notomi:2008}. The dielectric bands are characterized by field amplitudes maximized within the dielectric material, while the air bands exhibit large fields in the air holes. The dielectric-band modes can be confined by locally reducing the lattice constant, as the frequency of the modes will become higher and the photons will be surrounded by the photonic bandgap~\cite{Notomi:2008}. Similarly, air-band modes can be confined by introducing a larger lattice constant region~\cite{Zhang:2011,Miura:2014}. We expect that the air mode would have increased coupling with 2D materials because of stronger fields in air. Furthermore, the air band is closer to the light line than the dielectric band, corresponding to a lower effective refractive index with stronger evanescent fields. 

\begin{figure*}
\includegraphics{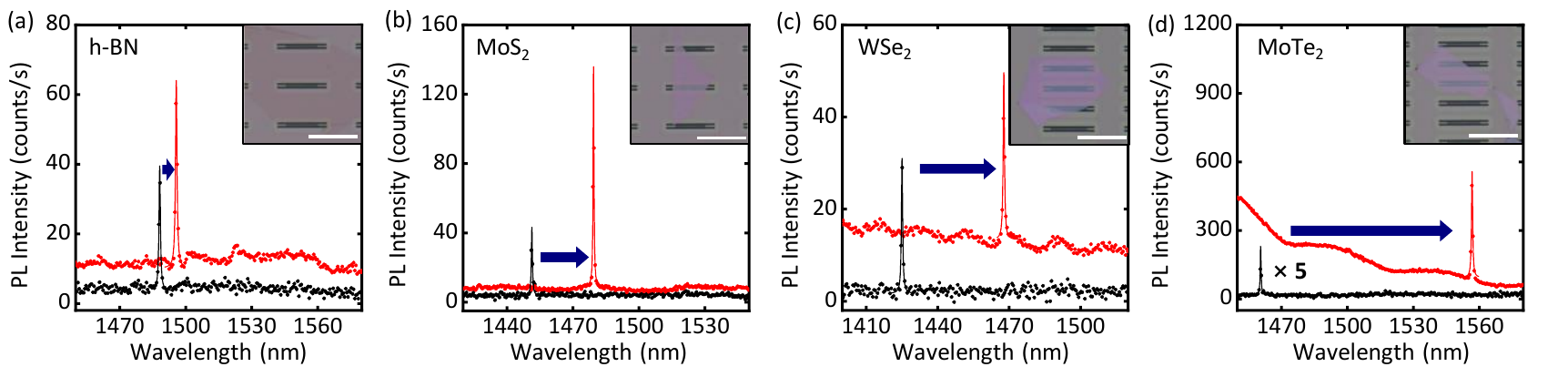}
\caption{
\label{Fig2} Resonant wavelength shifts by different types of 2D materials. PL spectra of an air-mode nanobeam before (black) and after (red) transfer of (a) a 12.5-nm-thick h-BN~flake, (b) a 11.0-nm-thick MoS$_2$~flake, (c) a 10.8 nm-thick~WSe$_2$ flake, and (d) a 11.9 nm-thick~MoTe$_2$ flake on top. The excitation power is 300~$\mu$W and the excitation wavelength is 780~nm. Arrows indicate the shift directions. Insets are optical images for the samples, and the scale bars are 20~$\mu$m. Dots are data and solid lines are Lorentzian peak fits.
}
\end{figure*}

The mode profiles in the two different cavity designs are mapped out by performing FDTD calculations. The $y$-component of the electric field $E_y$ for the fundamental transverse-electric mode in the dielectric-mode cavity is shown in Fig.~\ref{Fig1}c, and higher field amplitudes within silicon are confirmed as illustrated in the top view. In the air-mode cavity, the fields are mostly distributed within the holes as shown in Fig.~\ref{Fig1}d. The mode profiles are distinctly different not only in the $x$--$y$ plane but also for the evanescent fields in the $z$ direction as can be seen in the cross sectional views. The dielectric-mode cavity does show some evanescent fields, but the air-mode cavity exhibits even larger evanescent fields that extend farther out into free space as expected (see Supplementary Fig.~1). 

We experimentally compare the dielectric mode and the air mode cavities by transferring WSe$_2$ flakes onto the devices. The cavities are designed to have the fundamental mode in the telecommunication band based on the FDTD calculations, and we fabricate the nanobeams from a silicon-on-insulator substrate using electron beam lithography and inductively-coupled plasma etching. WSe$_2$ flakes with the same thickness of 9.0 nm are transferred on the two types of cavities using a conventional polydimethylsiloxane (PDMS) stamp method~\cite{CastellanosGomez:2014}. The interaction with the 2D material is evaluated with a home-built confocal photoluminescence (PL) microscopy system at room temperature by comparing the fundamental mode before and after the transfer of WSe$_2$. All measurements are performed in dry nitrogen to avoid environmental effects (see Supplementary Fig.~2).

PL spectra for the dielectric-mode and the air-mode cavities are shown in Fig.~\ref{Fig1}e and f, respectively. A single sharp peak is observed in each spectrum and is identified as the fundamental mode. The dielectric-mode cavity shows a redshift of $13.5$~nm after the transfer, which is attributed to a change in the average dielectric constant. In comparison, the air-mode cavity has an increased shift value by almost twofold. The larger shift indicates the enhanced responsivity, consistent with the simulations showing stronger fields for the air mode cavities. We note that higher-order modes show even more wavelength shifts because of the larger fraction of the mode volumes in free space (see Supplementary Fig.~3). 

We now survey the interaction of various 2D materials with the air-mode cavities. Four typical 2D materials are studied including insulating h-BN and semiconducting TMDs ranging from molybdenum disulfide (MoS$_2$), WSe$_2$, to molybdenum ditelluride (MoTe$_2$). We select flakes with thickness close to $11$~nm, and compare PL spectra before and after the transfer. As shown in Fig.~\ref{Fig2}, various 2D materials show distinct shift values because of the different dielectric constants. The smallest redshift of only $7.6$~nm is observed for h-BN, which is a widely studied 2D material for photonics~\cite{Wong:2015,Hayee:2020}.  For the TMDs, the redshifts increase in the order of MoS$_2$, WSe$_2$, and MoTe$_2$.  The tail of the MoTe$_2$ emission spectrum shows some overlap with the cavity mode, which is expected to lead to considerable cavity loss by absorption. In the following study, we make use of WSe$_2$ which exhibits large shifts with negligible emission in the present wavelength range~\cite{Kozawa:2014}.

\paragraph*{Giant wavelength tuning in WSe$_2$/cavity hybrids.}

\begin{figure}
\includegraphics {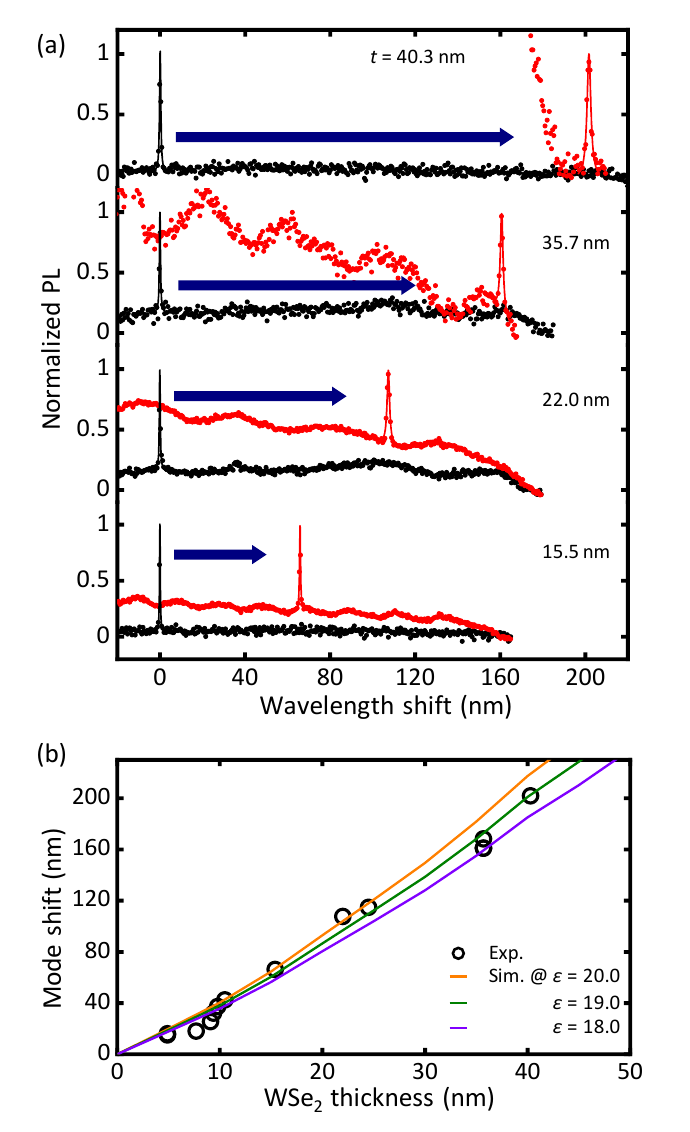}
\caption{
\label{Fig3} Thickness dependence of resonant wavelength shifts. (a) Normalized PL spectra of the fundamental mode for air-mode nanobeams before (red) and after (black) transfer of WSe$_2$ flakes. The data from top to bottom correspond to thickness of 40.3, 35.7, 22.0, 15.5~nm. Spectra are plotted with respect to the resonant wavelengths before the transfer, which are 1378.1, 1420.0, 1460.4, 1439.7~nm, respectively. The excitation power is 300~$\mu$W and the excitation wavelength is 780~nm. Arrows indicate the shift directions. Dots are data and solid lines are Lorentzian peak fits. (b) Resonant wavelength shifts as a function of WSe$_2$ thickness. Dots are data and lines are FDTD simulations with $\epsilon$ of 18.0 (purple), 19.0 (green), and 20.0 (orange). We note that several samples with the thin WSe$_2$ flakes deviate from the simulations, which could be due to the formation of an air space between the 2D material and the nanobeam (see Supplementary Fig.~4).
}
\end{figure}

To explore the capability of the enhanced responsiveness in WSe$_2$/cavity hybrids, modifications of the cavity mode by thick WSe$_2$ flakes are investigated. We transfer flakes with different thickness $t$ onto the cavities and mode shifts are measured by PL spectroscopy. As shown in Fig.~\ref{Fig3}a, the redshifts become larger with increasing $t$ owing to a larger mode overlap. The largest value of $201.8$~nm is achieved for $t=40.3$~nm, where the mode wavelength is close to the detection limit of our detector. We note that the tuning range covers a wide spectrum corresponding to the telecommunication E-band to L-band. 

We are able to extract the dielectric constant~$\epsilon$ from the thickness dependent shifts by comparing to  FDTD simulations (Fig.~\ref{Fig3}b). The calculated shifts are almost linearly dependent on the thickness when $t<10.0$~nm, and become superlinear in the thicker region which may be due to changes in the mode profile. The experimental data is reproduced well when $\epsilon=19\pm1$, whose value corresonds to the dielectric constant of bulk WSe$_2$. We note that in-plane $\epsilon$ is important for determining the redshifts because the fundamental mode is transverse electric (see Supplementary Fig.~5).

We also study the quality factor $Q$ in our WSe$_2$/cavity hybrids since large wavelength shifts are often accompanied by increased optical losses. As the spectral resolution of the present PL spectroscopy system is insufficient, transmittance measurements are performed to determine the $Q$ factor~\cite{McCutcheon:2011,Tetsumoto:2015}. For $t=22.0$~nm sample, the resonant wavelength is shifted by over $110$~nm while the $Q$ factor remains almost unaltered at $2.2\times10^4$ (see Supplementary Fig.~6). The negligible effect on the $Q$ factor could be attributed to a homogeneous interface with suppressed scattering loss as well as the absence of absorption from the WSe$_2$ flake at the cavity wavelength.

\paragraph*{Quantized wavelength shifts by atomically thin WSe$_2$.}

\begin{figure*}
\includegraphics{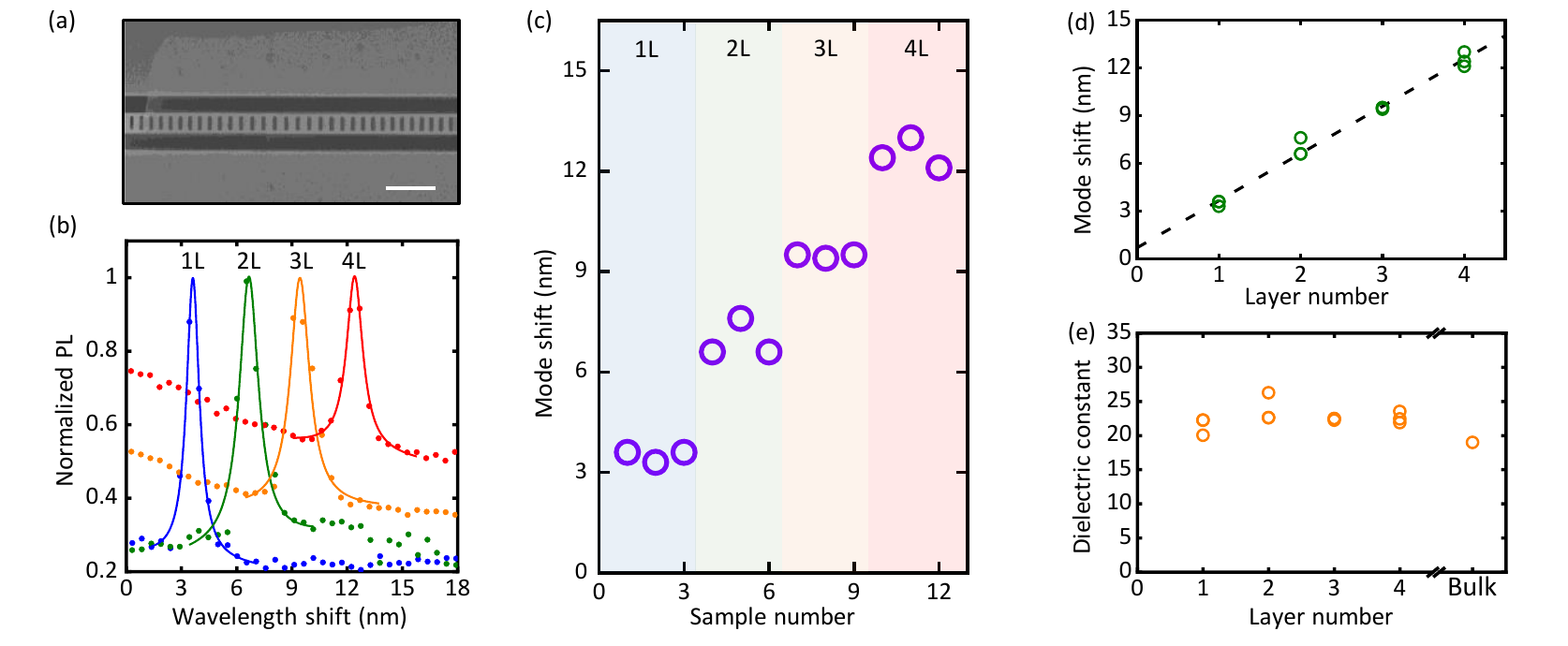}
\caption{
\label{Fig4} Quantized wavelength shifts induced by atomically thin WSe$_2$ flakes. (a) A top-view SEM image of a nanobeam cavity with a monolayer WSe$_2$ flake transferred above. Scale bar is 2~$\mu$m. (b) Normalized PL spectra of the fundamental modes for air-mode nanobeams with a monolayer (blue), bilayer (green), trilayer (orange), and quadlayer WSe$_2$ flake (red) on top. The horizontal axis is the shift with respect to the resonant wavelength before the transfer, which are 1461.6, 1454.9, 1464.2, 1468.4~nm, respectively. The excitation power is 300~$\mu$W and the excitation wavelength is 780~nm.  Dots are data and solid lines are Lorentzian plus linear fits. (c) Resonant wavelength shifts for all the measured samples. (d) Resonant wavelength shifts as a function of layer number. Dots are data and the broken line is a linear fit. (e) Extracted dielectric constant as a function of the layer number. Thickness of each layer is taken to be 0.65~nm in the analysis.
}
\end{figure*}

With the enhanced interaction, even a minor perturbation by 2D materials would influence the cavity mode and become detectable. By controlling the thickness of WSe$_2$ flakes with atomic precision, we demonstrate extreme sensitivity down to the monolayer limit. Since ultra-thin 2D materials are too fragile for sustaining the strain induced by PMDS stamps, deformation including rupture often occurs in the suspended region~\cite{Fryett:2018,Onodera:2020}. Here we utilize the anthracene-assisted transfer method~\cite{Otsuka:2021}, where the organic molecule single crystal supports and protects 2D flakes during the process (see Supplementary Fig.~7-9). Figure~\ref{Fig4}a shows a scanning electron microscopy (SEM) image of the prepared monolayer WSe$_2$/cavity hybrid, where a uniform morphology over the nanobeam can be observed.

The typical PL spectra of nanobeam cavities that are integrated with quad- (4L), tri- (3L), bi- (2L),  and mono-layer (1L) WSe$_2$ flakes are shown in Fig.~\ref{Fig4}b. The wavelength shift decreases by $\sim$3~nm as the thickness is reduced layer by layer. It is notable that the 1L WSe$_2$, the limit of the scaling, still tunes the mode of the nanobeam cavity effectively. The observed shifts are  almost two orders of magnitude larger than the linewidth, assuming $Q\sim10^4$. 

The roughly constant spacings between the cavity resonances suggest that the mode shifts are quantized due to the discrete thickness of atomically thin layers. To verify that the mode shifts are indeed governed by the layer number, we have prepared three sets of samples for each thickness and repeated the measurements (Fig.~\ref{Fig4}c). Clearly resolved steps are observed, confirming the quantization of the mode shifts. It is remarkable that the reproducible resonant shifts are obtained, benefitting from atomically precise thickness of the 2D material over the entire cavity. 

The observed discrete wavelength shifts offer an opportunity to study the dielectric constant even in the ultra-thin limit. We compare experimental results with FDTD simulation to extract $\epsilon$ for 1L to 4L WSe$_2$. Since calculated wavelength shifts are linearly dependent on $t$ in the thin limit (Fig.~\ref{Fig3}b), we fit the experimental results with a simple linear model (Fig.~\ref{Fig4}d). We observe a small intercept of 0.7~nm, which may come from anthracene residues as well as adsorption of other molecules on the surface of the 2D material. We have also performed FDTD simulations for various $\epsilon$, and find that wavelength shift $\Delta\lambda$ is also linearly dependent on $\epsilon$ (see Supplementary Fig.~5). The data can therefore be described by
\begin{equation}
\label{equation1}
\Delta\lambda = \alpha(\epsilon-1) t+\lambda_0,
\end{equation}
where $\alpha=0.21$ is the coefficient determined by the FDTD calculations and $\lambda_0$ is the intercept. Considering that anthracene residue or surface adsorption should occur similarly in all the samples, we take $\lambda_0=0.7$~nm. The values of $\epsilon$ are extracted from data using eq.~(\ref{equation1}), and the results are shown in Fig.~\ref{Fig4}e. The obtained values are comparable with the bulk value, revealing that the dielectric screening in the atomically thin limit is as effective. 

Since the cavity resonances arise from transverse-electric modes, the extracted values correspond to the in-plane $\epsilon$. It is reported that in-plane $\epsilon$ measured by spectroscopic ellipsometry decreases with increasing the layer number for ultra-thin TMDs~\cite{Yu:2015,Song:2019}. In comparison, first-principle calculations show that in-plane $\epsilon$ is insensitive to the layer number since the main contribution comes from in-plane atomic displacements~\cite{Laturia:2018}. Our results show no apparent thickness dependence down to the monolayer limit, which is consistent with the calculations. The behavior of in-plane $\epsilon$ is important for 2D materials based electronic devices, for example, the tunneling length through Schottky barriers in metal/2D contacts is strongly affected by in-plane $\epsilon$~\cite{Schulman:2018}.  We note that extrinsic effects from processing and adsorption have been taken into account through $\lambda_0$ in our analysis, and the extracted values should therefore represent the intrisic dielectric properties of atomically thin WSe$_2$.

\paragraph*{Flexible reconfiguration of 2D/cavity hybrids.}

\begin{figure}
\includegraphics{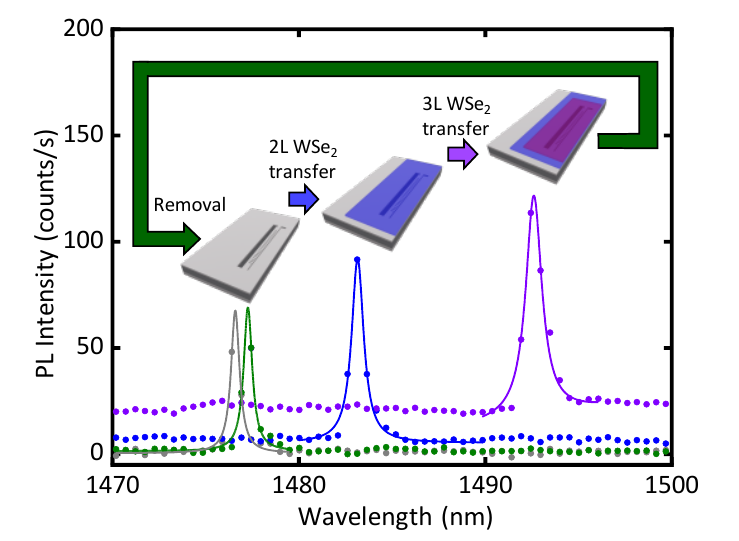}
\caption{
\label{Fig5} Reconfiguration of a cavity by stacking and removing of WSe$_2$ flakes. PL spectra of the fundamental mode of the as-fabricated cavity (grey), after the first transfer of a 2L WSe$_2$ flake (blue), after the second transfer of a 3L flake (purple), and after the removal of the stacked flakes (green). The excitation power is 300~$\mu$W and the excitation wavelength is 780~nm. Dots are data and solid lines are Lorentzian plus linear fits. Insets illustrate the configuration of the device during the measurements.
}
\end{figure}

One of the advantages of using 2D materials comes from their versatile manipulation capabilities, where flakes can be placed in the desired order without the restriction of lattice matching. In addition, the removal of transferred flakes is possible owing to the weak van der Waals force. In Fig.~\ref{Fig5}, we demonstrate additive and reversible control with a WSe$_2$/cavity hybrid. A 2L WSe$_2$ flake is first transferred on top of a nanobeam, and we observe a mode shift by $6.5$~nm (blue curve). Next, another transfer of a 3L flake is performed, which further shifts the mode by $9.5$~nm (purple curve). The shifts are consistent with the quantized values shown in Fig.~\ref{Fig4}c. We note that the linewidth increases slightly, suggesting additional loss from scattering induced by 2D-2D interfacial bubbles. By picking up the stack using an anthracene stamp (see Supplementary Fig.~8), the $Q$ factor is recovered and the mode is shifted back.

The high compatibility between 2D materials and cavities demonstrated here opens up various applications in photonics. The integration with 2D materials could be used as a post-fabrication tuning technology, where the giant wavelength tunable range of over $200$~nm can cover the telecommunication E-band through L-band. By using ultra-thin flakes, fine tuning can be done with high predictability and reproducibility because of the quantized shifts arising from the atomically precise thickness. This is in contrast to continuous mode shifts obtained in conventional methods such as chemical etching~\cite{Kuwabara:2019}, coating of photochromic films~\cite{Sridharan:2010}, and condensation of gases~\cite{Mosor:2005}. In addition, the stackable and removable 2D materials offer a new degree of freedom for reconfigurable cavities. Finally, as demonstrated in the measurement of $\epsilon$, 2D/cavity hybrids are an ideal platform for investigating the optical properties of atomically thin structures due to the high sensitivity of the air-mode cavities. Sensing of a small number of molecules could become possible by monitoring a high-$Q$ cavity resonance, which would extend the limits of the 2D materials based sensing technologies~\cite{Mao:2017}.

In conclusion, we have developed an air-mode cavity for integration with 2D materials. By investigating various 2D material types and thickness dependence, we show that the fabricated 2D/cavity hybrids have a giant wavelength tunable range of over $200$~nm. We observe a clear quantization behavior of the wavelength shift for WSe$_2$ where each step represents an effect from an additional monolayer. The in-plane dielectric constant of WSe$_2$ can be extracted because of the high surface sensitivity of the air-mode cavity, and shows a thickness independent characteristic down to the monolayer. By stacking and removing the transferred flakes, we demonstrate that the hybrids benefit from the versatile manipulation capabilities of the 2D materials. These findings reveal fascinating features of the cavity that is efficiently controlled by 2D materials, and provide a universal design strategy for enhancing the light-matter interaction with nanomaterials.

\section*{Methods}
\paragraph*{Nanobeam cavity fabrication.}
The nanobeam cavities are fabricated on a silicon-on-insulator substrate with a 260-nm-thick top silicon layer and a 1-$\mu$m-thick buried-oxide (BOX) layer. After defining the PC pattern on a resist mask by electron beam lithography, the pattern is transferred to the top silicon slab through an inductively coupled plasma process using C$_4$F$_8$ and SF$_6$ gases. Following resist removal, the BOX layer is etched with 20\% hydrofluoric acid to form an air-suspended nanobeam structure.

\paragraph*{PL measurements.}
A homebuilt confocal microscopy system is used to perform PL measurements at room temperature in dry nitrogen gas. We use a wavelength-tunable Ti:sapphire laser for excitation with its power controlled by neutral density filters. Polarization angle is adjusted to match the cavity mode by a half-wave plate. The laser beam is focused on the samples using an objective lens with a numerical aperture of 0.65 and a working distance of 4.5~mm. The $1/e^2$ beam diameter and the collection spot size defined by a confocal pinhole are $\sim$1.2 and 5.4~$\mu$m, respectively. PL is collected through the same objective lens and detected using a liquid-nitrogen-cooled 1024 pixel InGaAs diode array attached to a spectrometer. PL is dispersed by a 150 lines/mm grating and the dispersion is 0.52~nm/pixel at the center wavelength of 1340~nm. 

\paragraph*{Transfer of 2D materials by PDMS stamps.}
We transfer thick flakes of 2D materials with $t>5$~nm on the cavities by a conventional PDMS stamp method to form 2D/cavity hybrids~\cite{CastellanosGomez:2014}. 2D material flakes are prepared on a PDMS sheet (Gelfilm by Gelpak) by mechanical exfoliation and then transferred on the target nanobeam cavity at 120$^\circ$C by using a micromanipulator system. The thickness of the transferred 2D material flakes is determined by atomic force microscopy. h-BN crystals are supplied by NIMS and other 2D material crystals are purchased from HQ graphene.

\paragraph*{Transfer of 2D materials by anthracene crystals.}
Anthracene crystals are grown with the same method as in Ref.~\cite{Otsuka:2021}. 2D flakes are prepared on standard 90-nm-thick SiO$_2$/Si substrates by mechanical exfoliation and the layer number is determined by optical contrast. An anthracene single crystal is picked up with a glass-supported PDMS sheet to form an anthracene/PDMS stamp. 2D flakes are picked up by pressing the anthracene/PDMS stamp against a substrate with the target 2D flakes, followed by quick separation ($>10$~mm/s) so that the anthracene crystal remains attached to the PDMS sheet. The stamp is then pressed on a receiving substrate with the cavity. By slowly peeling off the PDMS ($<0.2$~$\mu$m/s), the anthracene crystal with the target 2D flake is released. Sublimation of anthracene in air at 110$^\circ$C for 10~min leaves behind clean 2D flakes because contamination from solvents is absent in the all-dry process.

\begin{acknowledgments}
Parts of this study are supported by JSPS (KAKENHI JP20H02558, JP20K15199, JP20J00817, JP20K15137, JP20K15120, JP19K23593, JP18H03864, JP19H00755, JP21H05237, JP21H05232), MIC (SCOPE 191503001), and MEXT (Nanotechnology Platform JPMXP09F19UT0075). The growth of hBN crystals is supported by the Elemental Strategy
Initiative conducted by the MEXT, Japan (Grant Number JPMXP0112101001)
and  JSPS KAKENHI (Grant Numbers JP19H05790, JP20H00354 and JP21H05233). D.Y. are supported by JSPS (Research Fellowship for Young Scientists). N.F. and S.F. are supported by RIKEN Special Postdoctoral Researcher Program. We thank the Advanced Manufacturing Support Team at RIKEN for technical assistance. We also thank H. Kumazaki for helpful discussion of the dimpled fiber fabrication.
\end{acknowledgments}

\section*{Author Contributions}
N.F. performed transfer of 2D materials as well as measurements of 2D materials and 2D/cavity hybrids. D.Y. performed simulation and fabrication of nanobeam cavities as well as measurements of 2D/cavity hybrids. S.F. developed the transmittance measurement system for 2D/cavity hybrids. K.O. helped develop anthracene-assisted dry method. T.T. and K.W. provided h-BN crystals. K.N. helped measurements of 2D materials. Y.K.K. supervised the project. N.F., D.Y., and Y.K.K. co-wrote the manuscript. All the authors commented on the manuscript. N.F. and D.Y. contributed equally to this work.

\end{document}